\documentclass{siamltex}  
\usepackage{amsmath}   
\usepackage{amssymb}

\input psfig.sty    

\newcommand{\pl}[2]{\frac{\partial#1}{\partial#2}}

\newcommand{\p}{\partial}    
\newcommand{\og}{\omega}    
\newcommand{\Og}{\Omega}    
\newcommand{\fl}[2]{\frac{#1}{#2}}    
\newcommand{\dt}{\delta}    
\newcommand{\tm}{\times}    
    
\newcommand{\nn}{\nonumber}    
\newcommand{\ap}{\alpha}    
\newcommand{\bt}{\beta}

\newcommand{\gm}{\gamma}

\newcommand{\ift}{\infty}

\newcommand{\sg}{\sigma}

\renewcommand{\theequation}{\arabic{section}.\arabic{equation}}    
\newcommand{\be}{\begin{equation}}    
\newcommand{\ee}{\end{equation}}    
\newcommand{\ba}{\begin{array}}    
\newcommand{\ea}{\end{array}}    
\newcommand{\bea}{\begin{eqnarray}}    
\newcommand{\eea}{\end{eqnarray}}    
\newcommand{\beas}{\begin{eqnarray*}}    
\newcommand{\eeas}{\end{eqnarray*}}    
    
\newtheorem{remark}{Remark}[section]

 \newcommand{\bx}{{\bf x} }

\begin{document}    
 
%\null\vskip -.5in 
%{\bf Submitted to} {\it SIAM J. Sci. Comput.}, Oct. 2003. 
%\medskip   
\title   
{A fourth-order time-splitting Laguerre-Hermite  
pseudo-spectral method  
for    Bose-Einstein condensates}    
\author{ Weizhu Bao    
\thanks{Department of Computational Science,   
National University of Singapore, Singapore 117543   
({\it bao@cz3.nus.edu.sg}). URL: http://www.cz3.nus.edu.sg/\~{}bao/. 
 Research is  supported  by the National University of Singapore    
 grant No. R-151-000-027-112.}   
 \and  Jie Shen    
\thanks{Department of Mathematics, Purdue University,  
West Lafayette, IN 47907, USA, ({\it shen@math.purdue.edu}). 
 Research is  supported  by   
NSF DMS-0311915 and he acknowledges support by 
the OAP (Fellow-Inbound) Programme supported by A*STAR and    
National University of Singapore.} }    
   
\date{}    
\maketitle   
   
\begin{abstract}    
A fourth-order     
time-splitting Laguerre-Hermite pseudospectral method is introduced for    
 Bose-Einstein condensates (BEC) in 3-D with cylindrical symmetry.   
 The method is   
  explicit, unconditionally stable, time reversible  and time transverse   
invariant. It conserves the position density, and is   
spectral accurate in space and fourth-order accurate   
in time. Moreover, the new method has two other important   
advantages:   
(i) it reduces a 3-D problem with cylindrical symmetry to an effective   
2-D problem;   
(ii) it solves the problem in the whole space instead of in a   
truncated artificial computational domain.   
The method is  applied to vector Gross-Pitaevskii equations    
(VGPEs)   
for multi-component BECs. Extensive numerical    
tests are  presented for 1-D GPE, 2-D GPE with radial symmetry,   
3-D GPE with cylindrical symmetry as well as   
3-D VGPEs for two-component BECs   
to show the efficiency and accuracy   
 of the new numerical method.   
\end{abstract}    
 
   
  
\pagestyle{myheadings} 
\markboth{W. Bao and J. Shen }{Laguerre-Hermite pseudospectral 
method for BEC}

\section{Introduction}\label{si}    
\setcounter{equation}{0}    
    
 Since its realization in dilute bosonic    
atomic gases \cite{Anderson,Bradley},    
Bose-Einstein    
condensation of alkali atoms and hydrogen has   
been produced and studied extensively in the laboratory \cite{Hall},    
and has  spurred great excitement in the    
atomic physics community and renewed the interest in studying the    
collective dynamics of macroscopic ensembles of atoms occupying the same    
one-particle quantum state \cite{General,Stringari,Griffin}.   
%The condensate typically consists of a few thousands    
% to millions of atoms which are confined    
%by a trap potential. In fact, beside the effects of the  
%internal interactions    
%between the atoms, the macroscopic behavior of BEC matter is highly    
%sensitive to the shape of this external trapping potential.    
Theoretical    
predictions of the properties of a BEC like the density profile    
\cite{Baym}, collective excitations \cite{colltheo} and the formation    
of vortices \cite{vortextheo} can now be compared with experimental data    
\cite{Anderson}. Needless to say that this dramatic    
progress on    
the experimental front has stimulated a wave of activity on both the    
theoretical and the numerical front.    
   
The properties of a BEC at temperatures $T$ much smaller than the    
critical condensation    
temperature $T_c$ \cite{LL} are usually well modeled by a    
nonlinear Schr\"{o}dinger equation (NLSE),  also called    
Gross-Pitaevskii equation (GPE) \cite{LL,Pit},   
 for the macroscopic wave function    
which incorporates the trap potential as well as the interactions among the    
atoms. The effect of the interactions is described by a mean field which    
leads to a nonlinear term in the GPE. The cases of repulsive and attractive    
interactions - which can both be realized in the experiment - correspond    
to defocusing and focusing nonlinearities in the GPE, respectively.   
 The results    
obtained by solving the GPE showed excellent agreement with most    
of the experiments (for a review see    
\cite{Anglin,Cornell}). In fact, up to now there have been    
very few experiments in ultra-cold dilute bosonic gases which could    
not be described properly by using theoretical methods based on    
the GPE \cite{NGPEExp,NGPETheo}. Thus developing efficient numerical    
methods for solving GPE is very important in numerical simulation    
of BEC.   
   
Recently,    a series of  numerical studies are devoted to   
 the numerical solution of  time-independent GPE for finding the   
 ground states    
and of time-dependent GPE for determining the dynamics of  BECs.    
To compute ground states of BECs, Bao and Du \cite{BD} presented   
a continuous normalized gradient flow (CNGF) with  diminishing energy,   
 and discretized it by a backward Euler finite difference (BEFD) method;   
Bao and Tang \cite{Bao} proposed a method which can be used    
to compute the ground and excited states via directly minimizing    
the energy functional; Edwards and Burnett \cite{Edwards} introduced    
a Runge-Kutta type method; other methods include   
an explicit imaginary-time algorithm    
 in \cite{Adu,Tosi}; a directly inversion in the iterated    
subspace (DIIS) in \cite{Feder}   
and a simple analytical type method in \cite{Dodd}.   
To determine the    
 dynamics of BECs, Bao et al. \cite{Bao3,baom,Bao4} presented    
a time-splitting spectral (TSSP) method, Ruprecht et al. \cite{Rup}   
used the Crank-Nicolson finite difference (CNFD) method, Cerimele    
et al. \cite{Tosi2,Cerim}   
proposed a particle-inspired scheme.    
   
 In most experiments of BECs, the magnetic trap is with cylindrical    
symmetry. Thus, the 3-D GPE in  Cartesian coordinate    
can be reduced to an effective 2-D problem    
in cylindrical coordinate. In this case, both the TSSP \cite{Bao3,Bao4,baom}   
and CNFD \cite{Rup} methods have serious drawbacks:   
(i) One needs to replace the original whole space    
by a truncated computational domain with an artificial (usually  homogeneous   
Dirichlet boundary conditions are used)  boundary condition.   
 How to choose an appropriate   
bounded computational domain  is a   
difficult task in practice:   
if it is too large, the computational resource is wasted;   
if it is too small, boundary effect will lead to wrong numerical    
solutions. (ii) The TSSP method is explicit and of spectral    
accuracy in space, but one needs to solve the original 3-D problem due    
to the periodic/homogeneous Dirichlet boundary conditions    
required by Fourier/sine spectral method. Thus, the memory    
requirement is a big burden in this case. The CNFD discretizes the    
2-D effective problem directly, but it is implicit and only second-order    
accurate in space.  The aim of this paper is to develop a numerical   
method which enjoys   
advantages of both TSSP and CNFD. That is to say,   
the method is explicit and of spectral   
order accuracy in space,  and discretizes the effective 2-D  
problem directly.     
We shall present such an efficient and accurate   
numerical method for discretizing    
 3-D GPE with cylindrical symmetry by applying a time-splitting technique   
and constructing appropriately scaled   
 Laguerre-Hermite  basis functions.

The paper is organized as follows. In Section \ref{sgpe}, we    
present the Gross-Pitaevskii equation and its dimension reduction.    
 In Section \ref{smethod},   
we present time-splitting Hermite, Laguerre and    
Laguerre-Hermite spectral methods for 1-D GPE,   
2-D GPE with radial symmetry and 3-D GPE with cylindrical    
symmetry, respectively. Extension of the time-splitting    
Laguerre-Hermite spectral method for vector Gross-Pitaevskii   
equations (VGPEs) for multi-component BEC is presented 
in Section \ref{exten}.  
In Section \ref{sne}, numerical results for 1-D GPE, 2-D GPE with   
radial symmetry, 3-D GPE with cylindrical symmetry as   
well as 3-D VGPEs for multi-component BEC are reported    
to demonstrate the efficiency and accuracy of our new numerical methods.   
Some concluding remarks are given in Section \ref{sc}.

 \section{The Gross-Pitaevskii equation  (GPE)}   
\label{sgpe}    
\setcounter{equation}{0}    
    
 At temperatures $T$ much smaller than the critical temperature    
$T_c$ \cite{LL}, a BEC is well described by the macroscopic wave function    
 $\psi = \psi({\bf x},t)$ whose evolution is governed by    
a self-consistent, mean field nonlinear Schr\"{o}dinger equation    
(NLSE) known as the  Gross-Pitaevskii equation (GPE)    
\cite{Gross,Pit}    
\begin{equation}    
\label{gpe1}    
i\hbar\frac{\partial \psi(\bx,t)}{\partial t}=    
-\frac{\hbar^2}{2m}\nabla^2 \psi(\bx,t)+    
V(\bx)\psi(\bx,t) +N U_0 |\psi(\bx,t)|^2\psi(\bx,t),    
\end{equation}    
where $m$ is the atomic mass, $\hbar$ is the Planck constant,    
$N$ is the number of atoms in the condensate,    
$V(\bx)$ is an external trapping potential.    
When  a harmonic trap potential is considered,    
$V(\bx)=\frac{m}{2}\left(\omega_x^2 x^2+\omega_y^2 y^2    
+\omega_z^2 z^2\right)$ with $\omega_x$, $\omega_y$ and    
$\omega_z$ being the trap    
frequencies in $x$, $y$ and $z$-direction, respectively.    
%For the following we assume (w.r.o.g.) $\omega_x\le\omega_y\le\omega_z$.    
%When $\omega_x=\omega_y=\omega_z$, the trap potential is isotropic.    
In most current BEC experiments, the traps are cylindrically symmetric,    
i.e. $\omega_x=\omega_y$.    
$U_0=4\pi \hbar^2 a_s/m$ describes the interaction between atoms in the    
condensate    
with the $s$-wave scattering length  $a_s$ (positive for repulsive    
interaction and negative for attractive interaction). It is convenient   
to normalize  the wave function by requiring   
 \begin{equation}    
\label{norm}    
\int_{{\Bbb R}^3} \; |\psi(\bx,t)|^2\;d\bx=1.    
\end{equation}    
    
\subsection{Dimensionless GPE}    
    
In order to scale the Eq.    
(\ref{gpe1}) under the normalization    
(\ref{norm}),  we introduce    
\begin{equation}    
\label{dml}    
\tilde{t}=\omega_m t, \quad \tilde{\bx}=\frac{\bx}{a_0}, \quad    
\tilde{\psi}(\tilde \bx,\tilde t)=a_0^{3/2} \psi(\bx,t),    
\quad \hbox{with}\quad  a_0=\sqrt{\hbar/m\omega_m},    
\end{equation}    
where $\og_m=\min\{\og_x,\og_y,\og_z\}$,    
$a_0$ is the length of the harmonic oscillator ground state.    
In fact,  we choose $1/\omega_m$ and $a_0$ as the dimensionless    
time and length units, respectively.    
 Plugging (\ref{dml}) into (\ref{gpe1}), multiplying  by    
$1/m \omega_m^2 a_0^{1/2}$,    
 and then removing all \~{}, we    
get the following dimensionless GPE    
under the normalization (\ref{norm}) in three dimension    
\begin{equation}    
\label{gpe2}    
i\;\frac{\partial \psi(\bx,t)}{\partial t}=-\frac{1}{2}\nabla^2    
\psi(\bx,t)+ V(\bx)\psi(\bx,t)    
+ \beta\; |\psi(\bx,t)|^2\psi(\bx,t),    
\end{equation}    
where $\beta=\frac{U_0 N}{a_0^3\hbar \omega_m}=\frac{4\pi a_sN}{a_0}$ and   
\[    
\label{poten}    
V(\bx)=\frac{1}{2}\left(\gm_x^2 x^2+\gamma_y^2 y^2+\gamma_z^2 z^2\right),\;   
{\rm with }\;\gamma_\ap=\frac{\omega_\ap}{\omega_m}\  (\ap=x,y,z).   
\]    
    
   There are two extreme regimes of the interaction parameter $\beta$:    
(1) $\beta=o(1)$, the Eq. (\ref{gpe2}) describes a    
weakly interacting  condensation;  (2)    
$\beta\gg1$, it corresponds to a strongly interacting condensation or    
to the semiclassical regime.    
   
There are two typical extreme regimes    
between the trap frequencies: (1) $\gm_x=1$, $\gamma_y\approx    
1$ and $\gamma_z\gg1$,    
it is a disk-shaped condensation; (2) $\gm_x\gg1$, $\gamma_y\gg 1$ and    
$\gamma_z=1$,    
it is a cigar-shaped condensation.    
In these two cases, the 3-D GPE (\ref{gpe2})    
can be approximately  reduced to a 2-D and  1-D equation respectively   
\cite{BeyondGPE,Bao3,Bao} as explained below.

\subsection{Reduction to lower dimension}

\noindent{\bf   Case I (disk-shaped condensation):}   
\[    
%\label{r2d}    
\og_x\approx \og_y, \quad \og_z\gg \og_x, \qquad \Longleftrightarrow    
\quad  \gm_x=1, \ \gm_y\approx1, \quad \gm_z\gg 1.    
\]    
Here, the 3-D GPE (\ref{gpe2}) can be reduced to 2-D  GPE  
with $\bx=(x,y)$ by    
assuming that the time evolution does not cause excitations along the    
$z$-axis, since  the excitations along the $z$-axis have large energy   
(of order     
$\hbar \og_z$)  compared to    
that along the $x$ and $y$-axis with energies of order   
$\hbar \og_x$.    
Thus, we may assume that the condensation wave function along    
the $z$-axis is always    
well described by the harmonic oscillator    
ground state wave function, and set    
\be    
\label{d2d}    
\psi=\psi_2(x,y,t)\phi_{\rm ho}(z) \qquad \hbox{with}\quad    
\phi_{\rm ho}(z)=(\gamma_z/\pi)^{1/4}\; e^{-\gamma_z z^2/2}.    
\ee    
Plugging (\ref{d2d}) into (\ref{gpe2}),  multiplying by    
$\phi^*_{\rm ho}(z)$ (where $f^*$ denotes the conjugate of a function $f$),    
integrating with respect to  $z$ over $(-\ift,\ift)$, we get    
\be    
\label{gpe2dd}    
i\;\frac{\partial \psi_2(\bx,t)}{\partial t}=-\fl{1}{2}\nabla^2 \psi_2+    
\fl{1}{2}\left(\gm_x^2 x^2+\gm_y^2 y^2+C\right) \psi_2    
+ \beta_2|\psi_2|^2\psi_2,    
\ee    
where    
\[\beta_2 = \beta \int_{-\ift}^\ift \phi_{\rm ho}^4(z)\,dz    
=\beta \sqrt{\frac{\gamma_z}{2\pi}}, \quad     
C=\int_{-\ift}^\ift \left(\gm_z^2 z^2|\phi_{\rm ho}(z)|^2+    
\left|\fl{d\phi_{\rm ho}}{dz}\right|^2\right)\;dz.\]    
Since this GPE is time-transverse invariant, we can replace    
$\psi_2\to \psi\;e^{-i\fl{C t}{2}}$ so that  the constant $C$ in    
the trap potential disappears,  and we obtain the  2-D effective  GPE:   
\be    
\label{gpe2d}    
i\;\pl{\psi(\bx,t)}{t}=-\fl{1}{2}\nabla^2 \psi+    
\fl{1}{2}\left(\gm_x^2 x^2+\gm_y^2 y^2\right) \psi    
+ \beta_2 |\psi|^2\psi.    
\ee    
Note that the observables, e.g. the position density $|\psi|^2$,  
 are not affected by dropping the constant $C$ in (\ref{gpe2dd}).

\noindent{\bf Case II (cigar-shaped condensation):}   
\[    
%\label{r2dd}    
\og_x\gg \og_z, \quad \og_y\gg \og_z \qquad \Longleftrightarrow \quad    
 \gm_x\gg1, \quad \gm_y\gg 1, \ \gm_z=1.   
\]    
Here, the 3-D GPE (\ref{gpe2}) can be reduced to a 1-D  GPE with $\bx=z$.    
Similarly as in the 2-D case, we can derive the following 1-D GPE   
\cite{BeyondGPE,Bao3,Bao}:   
\be    
\label{gpe1d}    
i\; \pl{\psi(z,t)}{t}=-\fl{1}{2}\psi_{zz}(z,t)+    
\fl{\gm_z^2 z^2}{2} \psi(z,t)    
+ \beta_1 |\psi(z,t)|^2\psi(z,t),    
\ee    
where $\beta_1 = \beta \sqrt{\gamma_x \gamma_y}/2\pi$.    
    
The 3-D GPE (\ref{gpe2}), 2-D GPE (\ref{gpe2d}) and    
1-D GPE (\ref{gpe1d}) can be written in a unified form:    
\bea    
\label{gpeg1}    
&&i\;\pl{\psi(\bx,t)}{t}=-\fl{1}{2}\nabla^2 \psi+    
V_d(\bx)\psi    
+ \beta_d\; |\psi|^2\psi, \quad \bx\in {\Bbb R}^d, \\   
\label{gpeg2}    
&&\psi(\bx,0)=\psi_0(\bx), \qquad \bx\in {\Bbb R}^d,   
\eea    
with   
\begin{equation*}   
\beta_d=\beta \left\{\ba{l}    
\sqrt{\gamma_x \gamma_y}/2\pi, \\    
\sqrt{\gamma_z/2\pi}, \\    
1,\\    
\ea\right.    
\quad    
V_d(\bx)=\left\{\ba{ll}    
 \gm_z^2 z^2/2,  &\quad d=1, \\    
\left(\gm_x^2 x^2+\gm_y^2 y^2\right)/2, &\quad d=2, \\    
\left(\gm_x^2 x^2+\gm_y^2 y^2+\gm_z^2 z^2\right)/2, &\quad d=3,\\    
\ea\right.    
\end{equation*}   
where $\gm_x>0$, $\gm_y>0$ and $\gm_z>0$ are  constants.    
The normalization condition for (\ref{gpeg1}) is    
\be    
\label{normg}    
N(\psi)=\|\psi(\cdot,t)\|^2=\int_{{\Bbb R}^d} \; |\psi(\bx,t)|^2\;d\bx   
\equiv \int_{{\Bbb R}^d} \; |\psi_0(\bx)|^2\;d\bx=1.    
\ee

\section{Fourth-order time-splitting  Laguerre-Hermite pseudospectral   
  method}    
\label{smethod}    
\setcounter{equation}{0}    
    
In this section we present a fourth-order time-splitting    
Laguerre-Hermite pseudospectral method for    
the problem (\ref{gpeg1})-(\ref{gpeg2}) in 3-D with cylindrical    
symmetry. As preparatory steps we begin by introducing    
the fourth-order time-splitting method and applying it with Hermite   
pseudospectral method for 1-D GPE    
and with Laguerre pseudospectral method for 2-D GPE with radial symmetry,   
respectively.     
   
Consider a general evolution equation   
\begin{equation}   
  \label{eq:gen}   
 i u_t=f(u)=Au+Bu 
\end{equation}   
where $f(u)$ is a nonlinear operator and the splitting $f(u)=Au+Bu$ 
 can be quite abitrary, in particular, $A$ and $B$ do not need to commute. 
For a given time step $\Delta t>0$, let   
$t_n=n\; \Delta t$,  $n=0,1,2,\ldots$ and $u^n$ be the approximation   
of $u(t_n)$.    
A fourth-order symplectic time integrator (cf. \cite{Yosh90,Lee})  for 
 (\ref{eq:gen}) 
  is as follows:   
\begin{equation}   
  \label{eq:split}   
  \begin{split}   
&u^{(1)}=e^{-i 2w_1A\Delta t}\;u^n;\\   
&u^{(2)}=e^{-i 2 w_2B\Delta t}\;u^{(1)};\\   
&u^{(3)}=e^{-i 2w_3A\Delta t}\;u^{(2)};\\   
&u^{(4)}=e^{-i 2w_4B\Delta t}\;u^{(3)};\\   
&u^{(5)}=e^{-i 2w_3A\Delta t}\;u^{(4)};\\       
&u^{(6)}=e^{-i 2w_2B\Delta t}\;u^{(5)};\\   
&u^{n+1}=e^{-i 2w_1A\Delta t}\;u^{(6)};\\   
  \end{split}   
\end{equation}   
where   
\begin{equation}   
  \label{eq:wvalues}   
  \begin{split}   
&w_1=0.33780\;17979\;89914\;40851,\;   
w_2 = 0.67560\;35959\;79828\;81702,\\   
&w_3=-0.08780\;17979\;89914\;40851,\;   
w_4  =-0.85120\;71979\;59657\;63405.   
  \end{split}   
\end{equation}   
   
We now rewrite the GPE (\ref{gpeg1}) in the form of   
(\ref{eq:gen}) with    
\be   
A\psi=\beta_d\; |\psi(\bx,t)|^2\psi(\bx,t),\qquad   
B\psi=-\fl{1}{2}\nabla^2 \psi(\bx,t)+    
V_d(\bx)\psi(\bx,t).   
\ee   
Thus, the key for an efficient implementation of (\ref{eq:split})    
is to solve efficiently the following two subproblems:   
\be   
\label{ode}   
i\;\pl{\psi(\bx,t)}{t}=A\psi(\bx,t)   
=\beta_d\; |\psi(\bx,t)|^2\psi(\bx,t),    
\qquad \bx\in {\Bbb R}^d,   
\ee   
and   
\be   
\label{step1}   
\begin{split}   
&i\;\pl{\psi(\bx,t)}{t}=B\psi(\bx,t)=-\fl{1}{2}\nabla^2 \psi(\bx,t)+    
V_d(\bx)\psi(\bx),  \qquad   \bx\in {\Bbb R}^d,\\   
&\lim_{|\bx|\rightarrow +\infty}\psi(\bx,t)=0.   
\end{split}   
\ee   
The decaying condition in (\ref{step1}) is necessary for   
satisfying the normalization (\ref{normg}).   
   
Multiplying (\ref{ode})  by $\overline{\psi(\bx,t)}$, we find that 
 the ordinary differential   
equation (\ref{ode})   
leaves $|\psi(\bx,t)|$ invariant in $t$. Hence,    
for $t\ge t_s$ ($t_s$ is any given time),  (\ref{ode}) becomes   
\be   
\label{ode1}   
i\;\pl{\psi(\bx,t)}{t}   
=\beta_d\; |\psi(\bx,t_s)|^2\psi(\bx,t), \;t\ge t_s,   
\qquad \bx\in {\Bbb R}^d   
\ee   
which can be integrated {\bf exactly}, i.e.,   
\be   
\label{sol}   
\psi(\bx,t)= e^{-i\beta_d |\psi(\bx,t_s)|^2 (t-t_s)}   
\psi(\bx,t_s),     
\qquad t\ge t_s, \quad \bx\in {\Bbb R}^d.   
\ee   
Thus, it remains to find an efficient and accurate scheme for (\ref{step1}).   
 We shall construct below suitable spectral basis functions which are   
eigenfunctions of $B$ so that    
 $e^{-i B \Delta t}\psi$ can be exactly evaluated (which is necessary  
 for the final scheme to be time reversible and time transverse invariant).  
Hence,  the only time discretization error of    
the corresponding time splitting  method (\ref{eq:split}) is the splitting    
error, which is fourth  order in $\Delta t$.  
Furthermore, the scheme  is explicit,   
time reversible and time transverse invariant, and as we shall show  
below, it is also  unconditionally stable.   
   
\subsection{Hermite pseudospectral method for the 1-D GPE}   
   
In the 1-D case, Eq. (\ref{step1}) collapses to   
\bea   
\label{gpeg11d}    
&&i\;\pl{\psi(z,t)}{t}=B\psi(z,t)=-\fl{1}{2}\fl{\p^2\psi(z,t)}{\p z^2}+    
\fl{\gm_z^2 z^2}{2}\psi(z,t),  \quad z\in {\Bbb R},\\   
&&\lim_{|z|\rightarrow +\infty}\psi(z,t)=0,\;t\ge 0,   
\eea   
with the normalization (\ref{normg})   
\be   
\|\psi(\cdot, t)\|^2=\int_{-\ift}^\ift |\psi(z,t)|^2  dz \equiv   
 \int_{-\ift}^\ift |\psi_0(z)|^2  dz=1.   
\ee   
Since the above equation is posed on the  whole line, it is   
natural to consider Hermite functions which  have been successfully applied   
to other equations (cf. \cite{Funa92,GSX03}). Although the   
standard Hermite functions could be used as basis functions here, they   
are not the most appropriate.  
 Below, we construct properly scaled    
 Hermite functions which are eigenfunctions of $B$.   
   
Let $H_l(z)$ $(l=0,1,\ldots,N)$ be  the standard Hermite polynomials   
satisfying   
\bea   
\label{herm1d}   
&&H_l^{\prime\prime}(z) -2 z H_l^\prime(z) +2l H_l(z)=0,    
\qquad z\in {\Bbb R}, \quad l\ge0,\\   
\label{herm2d}   
&&\int_{-\ift}^\ift H_l(z) H_n(z) e^{-z^2}\;dz =   
\sqrt{\pi}\; 2^l\; l!\; \dt_{ln}, \qquad l,n\ge0,   
\eea   
where $\dt_{ln}$ is the Kronecker delta.   
We define the scaled Hermite function   
\be   
\label{mherm}   
h_l(z)=e^{-\gm_z z^2/2}\, H_l\left(\sqrt{\gm_z} z\right)/   
\sqrt{2^l\, l!} (\pi/\gm_z)^{1/4}, \qquad  z\in {\Bbb R}.   
\ee   
 Plugging (\ref{mherm}) into    
(\ref{herm1d}) and (\ref{herm2d}), we find that    
\bea   
\label{herm1dd}   
&&-\fl{1}{2}h_l^{\prime\prime}(z) + \fl{\gm_z^2 z^2}{2} h_l(z)   
= \mu_l^zh_l(z),    
\ z\in {\Bbb R}, \qquad \mu_l^z =\fl{2l+1}{2}\gm_z, \quad l\ge0,\\   
\label{herm2dd}   
&&\int_{-\ift}^\ift h_l(z) h_n(z) \;dz =   
\int_{-\ift}^\ift    
\fl{1}{\sqrt{\pi2^l l!2^n n!}} H_l(z) H_n(z) e^{-z^2}\;dz=   
\dt_{ln}, \quad l,n\ge0.\qquad    
\eea   
Hence, $\{h_l\}$ are eigenfunctions of $B$  defined in 
(\ref{gpeg11d}).    
   
For a fixed $N$, let $ X_N={\rm span}\{h_l:\,l=0,1,\cdots,N\}$. The   
Hermite-spectral method for  (\ref{gpeg11d}) is    
 to find $\psi_N(z,t)\in X_N$, i.e.,   
\be   
\label{expan1d}   
\psi_N(z,t)=\sum_{l=0}^N \hat{\psi}_l(t) \; h_l(z), \qquad  z\in {\Bbb R}.   
\ee   
 such that   
\be    
\label{gpeg11dn}    
i\;\pl{\psi_N(z,t)}{t}=B\psi_N(z,t)=-\fl{1}{2}\fl{\p^2\psi_N(z,t)}{\p z^2}+    
\fl{\gm_z^2 z^2}{2}\psi_N(z,t),  \quad z\in {\Bbb R}.    
\ee    
Note that $\lim_{|z|\rightarrow +\infty}h_l(z)=0$ (cf. \cite{Szeg75})   
so the decaying condition $\lim_{|z|\rightarrow +\infty}\psi_N(z,t)$ $=0$  
is automatically    
satisfied.   
   
Plugging (\ref{expan1d}) into (\ref{gpeg11dn}), thanks to   
(\ref{herm1dd}) and   
(\ref{herm2dd}), we find   
\be   
\label{ode1d2}   
i\; \fl{d \hat{\psi}_l(t)}{dt}= \mu_l^z\;\hat{\psi}_l(t)=   
 \fl{2l+1}{2}\gm_z\; \hat{\psi}_l(t),\qquad l=0,1,\cdots,N.   
\ee   
Hence, the solution for  (\ref{gpeg11dn}) is given by   
 \be   
\label{sol1d3}   
\psi_N(z,t)=e^{-iB(t-t_s)}\psi_N(z,t_s)   
 = \sum_{l=0}^N e^{-i   
  \mu_l^z (t-t_s)}\; \hat{\psi}_l(t_s)h_l(z),    
\qquad t\ge t_s.   
\ee   
   
  Let $\{\hat{z}_k\}_{k=0}^N$ be the Hermite-Gauss points   
  (cf. \cite{Szeg75,Funa92}), i.e.     
$\{\hat{z}_k\}_{k=0}^N$ are the $N+1$ roots of the polynomial   
  $H_{N+1}(z)$.   
Let $\psi^n_k$ be the approximation of $\psi(z_k,t_n)$ and    
$\psi^n$ be the solution vector with components $\psi_k^n$.    
Then, the fourth-order time-splitting Hermite-pseudospectral (TSHP4) method    
for 1-D GPE (\ref{gpeg1}) is given by    
\bea    
&&\psi^{(1)}_k=e^{-i2w_1\; \Delta t\; \bt_1|\psi_k^{n}|^2}   
\;\psi_k^{n},    \nn\\    
&&\psi_k^{(2)}=\sum_{l=0}^{N}     
  e^{-i2w_2\; \mu_l^z \;\Delta t }\;   
\widehat{(\psi^{(1)})}_l\ h_l(z_k),      
    \nn\\     
\label{schm4}     
&&\psi^{(3)}_k=e^{-i2w_3\; \Delta t\;    
 \beta_1|\psi_k^{(2)}|^2}\;\psi_k^{(2)},   \nn\\    
&&\psi_k^{(4)}=\sum_{l=0}^{N}     
  e^{-i2w_4\; \mu_l^z \;\Delta t }\;   
\widehat{(\psi^{(3)})}_l\ h_l(z_k),     
    \qquad k=0,1,\ldots,N,\nn\\     
&&\psi^{(5)}_k=e^{-i2w_3 \; \Delta t\;    
 \beta_1|\psi_k^{(4)}|^2}\;\psi_k^{(4)},    \nn\\   
&&\psi_k^{(6)}=\sum_{l=0}^{N}     
  e^{-i2w_2\; \mu_l^z \;\Delta t }\;   
\widehat{(\psi^{(5)})}_l\ h_l(z_k),      \nn\\     
&&\psi^{n+1}_k=e^{-i2w_1 \; \Delta t\;   
 \beta_1|\psi_k^{(6)}|^2}\;\psi_k^{(6)};     
\eea    
where $w_i, \;i=1,2,3,4$ are given in (\ref{eq:wvalues}), and   
$\{\widehat{U}_l\}$, the coefficients of  scaled Hermite  
expansion of $U(z)$,    
can be computed from the discrete scaled Hermite transform:   
\be    
\label{Fourc1}    
\widehat{U}_l=    
  \sum_{k=0}^{N} \ \og_k^z\; U(z_k)\; h_l(z_k),     
\qquad l=0,1,\ldots, N.   
\ee    
In the above, $z_k$ and $\og_k^z$ are  
 the  scaled Hermite-Gauss points and weights, respectively,   
which are defined by   
\be    
\label{hweight}    
\og_k^z = \frac{\hat{\og}_k^z\;  e^{\hat{z}_k^2}}{\sqrt{\gm_z}},   
\qquad z_k= \frac{\hat{z}_k}{\sqrt{\gm_z}},    
\qquad 0\le k\le N,  
\ee  
where   
 $\{\hat{\og}_k^z\}_{k=0}^N$ are the  weights associated with the   
Hermite-Gauss quadrature (cf.  \cite{Funa92})    
satisfying    
\be   
\sum_{k=0}^{N}\ \hat{\og}_k^z \ \fl{H_l(\hat{z}_k)}   
{\pi^{1/4}\,\sqrt{2^l\, l!}}\   
\fl{H_n(\hat{z}_k)}{\pi^{1/4}\,\sqrt{2^n\, n!}} = \dt_{ln},    
\qquad l,n=0,1,\ldots,N,   
\ee   
and we derive from (\ref{mherm}) that   
\bea   
\sum_{k=0}^{N} \ \og_k^z\; h_l(z_k)\; h_m(z_k)   
&=&\sum_{k=0}^{N}\  \hat{\og}_k^z\,  e^{\hat{z}_k^2}/\sqrt{\gm_z}\    
h_l\left(\hat{z}_k/\sqrt{\gm_z} \right) \   
h_m\left(\hat{z}_k/\sqrt{\gm_z} \right) \nn\\   
&=&\sum_{k=0}^{N}\ \hat{\og}_k^z\ \fl{H_l(\hat{z}_k)}   
{\pi^{1/4}\sqrt{2^l\, l!}}\ \fl{H_n(\hat{z}_k)}{\pi^{1/4}\sqrt{2^n\,   
    n!}}=\delta_{ln},   
 \; 0\le l,n\le N.  \qquad  
\label{doherm}   
\eea   
 Note that the computation of $\{\omega^z_k\}$ from (\ref{hweight}) is  
 not a stable process for very large $N$. However, one can compute   
 $\{\omega^z_k\}$ in a stable way as suggested in the Appendix of  
 \cite{Shen00}.   
  
Thus, the memory requirement of this scheme is $O(N)$ and    
the computational cost per time step is a small multiple of $N^2$.   
As for the stability of the TSHP4, we have the following    
 
\begin{lemma} \label{tshp4sb}  
The time-splitting  Hermite-pseudospectral (TSHP4) method (\ref{schm4})   
is unconditionally stable. More precisely, we have   
\be   
\label{stab} 
\|\psi^n\|_{l^2}^2=\sum_{k=0}^N \og_k^z |\psi_k^n|^2   
= \sum_{k=0}^M \og_k^z |\psi_0(z_k)|^2=\|\psi_0\|_{l^2}^2, \qquad    
n=0,1,\ldots\;.   
\ee   
\end{lemma}   
\begin{proof} From (\ref{schm4}), noting (\ref{Fourc1}) and 
 (\ref{doherm}), we obtain 
\bea 
\label{parsval} 
\|\psi^{n+1}\|_{l^2}^2&=&\sum_{k=0}^N \og_k^z |\psi_k^n|^2   
= \sum_{k=0}^N \og_k^z \left|e^{-i2w_1 \; \Delta t\;   
 \beta_1|\psi_k^{(6)}|^2}\;\psi_k^{(6)}\right|^2\nn\\ 
&=&\sum_{k=0}^N \og_k^z |\psi_k^{(6)}|^2  
=\sum_{k=0}^N \og_k^z \left|\sum_{l=0}^{N}     
  e^{-i2w_2\; \mu_l^z \;\Delta t }\;   
\widehat{(\psi^{(5)})}_l\ h_l(z_k)\right|^2\nn\\ 
&=&\sum_{l=0}^{N} \sum_{m=0}^{N}   
e^{-i2w_2\; \mu_l^z \;\Delta t }\;   
\widehat{(\psi^{(5)})}_l e^{i2w_2\; \mu_m^z \;\Delta t }\;   
(\widehat{(\psi^{(5)})}_m)^*  \left[\sum_{k=0}^N \og_k^z h_l(z_k) h_m(z_k) 
\right]\nn\\ 
&=&\sum_{l=0}^{N} \sum_{m=0}^{N}   
e^{-i2w_2\; \mu_l^z \;\Delta t }\;   
\widehat{(\psi^{(5)})}_l e^{i2w_2\; \mu_m^z \;\Delta t }\;   
(\widehat{(\psi^{(5)})}_m)^*  \ \dt_{lm}\nn\\ 
&=&\sum_{l=0}^{N} |\widehat{(\psi^{(5)})}_l|^2 = 
\sum_{l=0}^{N} \left|\sum_{k=0}^{N} \ \og_k^z\; \psi^{(5)}(z_k)\;  
h_l(z_k)\right|^2\nn\\ 
&=&\sum_{k=0}^{N} \sum_{m=0}^{N} \og_k^z  \psi^{(5)}(z_k) \psi^{(5)}(z_m)^* 
\left[\sum_{l=0}^{N}\og_m^z h_l(z_k) h_l(z_m)\right]\nn\\ 
&=&\sum_{k=0}^{N} \sum_{m=0}^{N} \og_k^z  \psi^{(5)}(z_k) \psi^{(5)}(z_m)^* 
\ \dt_{km}\nn\\ 
&=&\sum_{k=0}^{N} \og_k^z |\psi^{(5)}(z_k)|^2 = \|\psi^{(5)}\|_{l^2}^2. 
\eea 
Similarly, we have  
\be 
\label{relat} 
\|\psi^{n+1}\|_{l^2}^2 = \|\psi^{(5)}\|_{l^2}^2 = \|\psi^{(3)}\|_{l^2}^2 
=\|\psi^{(1)}\|_{l^2}^2 = \|\psi^{n}\|_{l^2}^2, \qquad n\ge0. 
\ee 
Thus the equality (\ref{stab}) can be obtained from (\ref{relat}) by  
induction. 
%Using (\ref{doherm}),  the proof is essentially the same   
%  as in  \cite{Bao1,BJP} for the time-splitting     
%spectral method.   
\end{proof}   
   
\begin{remark}   
Extension of TSHP4 method (\ref{schm4}) to 2-D GPE without radial    
symmetry  and 3-D GPE without cylindrical symmetry is straightforward   
by using tensor product of scaled Hermite functions.   
\end{remark}

\subsection{Laguerre pseudospectral method for 2-D GPE with radial symmetry}   
   
In the 2-D case with radial symmetry, i.e. $d=2$ and $\gm_x=\gm_y$   
in (\ref{gpeg1}), and $\psi_0(x,y)=\psi_0(r)$ in (\ref{gpeg2})    
with $r=\sqrt{x^2+y^2}$, we can write    
the solution of (\ref{gpeg1}), (\ref{gpeg2}) as   
$\psi(x,y,t)=\psi(r,t)$. Therefore, Eq. (\ref{step1}) collapses to   
\bea    
\label{gpeg2d}    
&&i\;\pl{\psi(r,t)}{t}=B\psi(r,t)   
=-\fl{1}{2r}\pl{}{r}\left(r\pl{\psi(r,t)}{r}\right)+     
\fl{\gm_r^2 r^2}{2}\psi(r,t),  \quad 0<r<\ift,\\   
&&\lim_{r\to\ift} \psi(r,t)=0, \qquad t\ge0, \label{gpeg2d2}    
\eea    
where $\gm_r=\gm_x=\gm_y$. The normalization (\ref{normg})   
collapses to    
\be   
\label{norm5}   
\|\psi(\cdot, t)\|^2=2\pi \int_0^\ift |\psi(r,t)|^2 r\; dr \equiv   
2\pi \int_0^\ift |\psi_0(r)|^2 r\; dr=1.   
\ee   
Note that it can be shown, similarly as for the Poisson equation in a   
2-D disk (cf. \cite{Shen97}), that the problem   
(\ref{gpeg2d})-(\ref{gpeg2d2}) admits a unique solution without any   
condition at the pole $r=0$.    
   
Since (\ref{gpeg2d}) is posed on a semi-infinite interval, it is   
natural to consider  Laguerre functions which have been successfully used   
for other problems in semi-infinite intervals   
(cf. \cite{Funa92,Shen00}). Again, the standard Laguerre functions,   
although usable, are not the most appropriate for this problem. Below,   
we construct properly scaled Laguerre functions which are   
eigenfunctions of $B$.   
   
Let $\hat{L}_m(r)$ $(m=0,1,\ldots,M)$ be   
the  Laguerre polynomials of degree $m$    
satisfying   
\bea   
\label{lau2d}   
&&r\hat{L}_m^{\prime\prime}(r)+(1-r) \hat{L}_m^\prime (r) +m \hat{L}_m(r)=0,   
\qquad m=0,1,\ldots,\\   
\label{lau2d2}   
&&\int_0^\ift\; e^{-r}\; \hat{L}_m(r)\; \hat{L}_n(r)\;dr =\dt_{mn},   
\qquad m,n=0,1,\ldots\;.   
\eea   
We define the scaled    
Laguerre functions $L_m$ by   
\be   
\label{mlagu}   
L_m(r)=\sqrt{\frac{\gm_r}{\pi}}\ e^{-\gm_r r^2/2}\; \hat{L}_m   
(\gm_r r^2), \qquad 0\le r<\ift.   
\ee   
Note that $\lim_{|r|\rightarrow +\infty}L_m(r)=0$ (cf. \cite{Szeg75})   
hence,  $\lim_{|r|\rightarrow +\infty}\psi_M(r,t)=0$ is automatically   
satisfied.   
   
 Plugging (\ref{mlagu}) into    
(\ref{lau2d}) and (\ref{lau2d2}), a simple computation shows   
\bea   
\label{lau3d}   
&&-\fl{1}{2r}\pl{}{r}\left(r\pl{L_m(r)}{r}\right) +\fl{1}{2}   
\gm_r^2 r^2 L_m(r) = \mu_m^r L_m(r), \ \mu_m^r=\gm_r(2m+1), \    
m\ge0, \qquad \quad \\   
\label{lau4d}   
&&2\pi \int_0^\ift L_m(r) L_n(r)r \;dr =   
\int_0^\ift e^{-r} \hat{L}_m(r) \hat{L}_n(r)\;dr=   
\dt_{mn},   
\quad m,n\ge0.   
\eea   
Hence, $\{L_m\}$ are eigenfunctions of $B$  defined in 
(\ref{gpeg2d}).    
   
For a fixed $M$, let $Y_M=\text{span}\{L_m:\,m=0,1,\cdots,M\}$. The   
 Laguerre-spectral method for  (\ref{gpeg11d}) is    
 to find $\psi_M(r,t)\in Y_M$, i.e.,   
\be   
\label{expan1dm}   
\psi_M(r,t)=\sum_{m=0}^M \hat{\psi}_m(t) \; L_m(r), \qquad  0\le r<\infty   
\ee   
 such that   
\begin{equation}   
\label{gpeg2dm}    
i\;\pl{\psi_M(r,t)}{t}=B\psi_M(r,t)=-\fl{1}{2r}\pl{}{r}  
\left(r\pl{\psi_M(r,t)}{r}\right)+       
\fl{\gm_r^2 r^2}{2}\psi_M(r,t),  \quad 0<r<\ift.   
\end{equation}   
Plugging (\ref{expan1dm}) into (\ref{gpeg2dm}), thanks to   
(\ref{lau3d}) and   
(\ref{lau4d}), we find   
\be   
\label{ode1d4}   
i\; \fl{d \hat{\psi}_m(t)}{dt}= \mu_m^r \hat{\psi}_m(t)=\gm_z (2m+1)    
\hat{\psi}_m(t),\;m=0,1,\cdots,M.    
\ee   
Hence, the solution for  (\ref{gpeg2dm}) is given by   
\be   
\label{sol1d2}   
\psi_M(r,t)=e^{-iB(t-t_s)}\psi_M(r,t_s)   
 = \sum_{m=0}^M e^{-i   
  \mu_m^r (t-t_s)}\; \hat{\psi}_m(t_s)L_m(r),    
\qquad t\ge t_s.   
\ee

Let $\{\hat{r}_j\}_{j=0}^M$ be the Laguerre-Gauss-Radau points   
  (cf. \cite{Funa92}), i.e.     
they are the $M+1$ roots of the polynomial    
$r\hat{L}'_{M+1}(r)$.   
Let $\psi^n_j$ be the approximation of $\psi(r_j,t_n)$ and    
$\psi^n$ be the solution vector with components $\psi_j^n$.    
Then, the fourth-order time-splitting Laguerre-pseudospectral 
(TSLP4) method    
for 2-D GPE (\ref{gpeg1}) with radial symmetry   
 is given by    
\bea    
&&\psi^{(1)}_j=e^{-i2w_1\; \Delta t\;  \bt_2|\psi_j^{n}|^2}   
\;\psi_j^{n},    \nn\\    
&&\psi_j^{(2)}=\sum_{l=0}^{M}     
  e^{-i2w_2\; \mu_l^r\; \Delta t}\;\widehat{(\psi^{(1)})}_l\;L_l(r_j),      
    \nn\\     
\label{schm5}     
&&\psi^{(3)}_j=e^{-i2w_3 \; \Delta t\;\beta_2|\psi_j^{(2)}|^2}\; 
\psi_j^{(2)},     
 \nn\\    
&&\psi_j^{(4)}=\sum_{l=0}^{M}     
  e^{-i2w_4 \;\mu_l^r\; \Delta t  }\;\widehat{(\psi^{(3)})}_l\; L_l(r_j),      
    \qquad j=0,1,\ldots,M,\nn\\     
&&\psi^{(5)}_j=e^{-i2w_3 \; \Delta t\; \beta_2   
|\psi_j^{(4)}|^2}\;\psi_j^{(4)},    \nn\\   
&&\psi_j^{(6)}=\sum_{l=0}^{M}     
  e^{-i2w_2\; \mu_l^r\; \Delta t}\;\widehat{(\psi^{(5)})}_l\;L_l(r_j),     
    \nn\\     
&&\psi^{n+1}_j=e^{-i2w_1 \; \Delta t\;    
\beta_2|\psi_j^{(6)}|^2}\;\psi_j^{(6)};     
\eea    
where    
$\widehat{U}_l$, the  coefficients of  scaled Laguerre expansion of $U(r)$   
can be computed from the discrete scaled Laguerre transform:   
\be    
\label{Fourc2}    
\widehat{U}_l=   \sum_{j=0}^{M} \og_j^r\; U(r_j)\; L_l(r_j),     
\qquad l=0,1,\ldots, M.   
\ee    
In the above, $r_j$ and $\og_j^z$ are  
 the  scaled Laguerre-Gauss-Radau points and weights, respectively,   
which are defined by   
\be\label{lweight}  
\og_j^r =\frac{\pi}{\gm_r} \;\hat{\og}_j^r\; e^{\hat{r}_j },   
\qquad r_j=\sqrt{\fl{\hat{r}_j}{\gm_r}},    
\qquad j=0,1,\ldots,M,  
\ee  
where   
 $\{\hat{\og}_j^r\}_{j=0}^M$ are the  weights associated to the   
  Laguerre-Gauss quadrature \cite{Funa92}   
satisfying    
\[   
\sum_{j=0}^{M} \hat{\og}_j^r \hat{L}_m(\hat{r}_j)   
\hat{L}_n(\hat{r}_j)= \dt_{nm},    
\qquad n,m=0,1,\ldots,M,\]   
and we derive from   
 (\ref{mlagu}) that   
\bea   
\sum_{j=0}^{M} \og_j^r L_m(r_j) L_n(r_j)   
&=& \sum_{j=0}^{M}  \hat{\og}_j^r e^{\hat{r}_j }\pi/\gm_r\    
L_m\left(\sqrt{\hat{r}_j/\gm_r}\right) \    
L_n\left(\sqrt{\hat{r}_j/\gm_r}\right)\nn\\   
&=&\sum_{j=0}^{M} \hat{\og}_j^r \hat{L}_m(\hat{r}_j)   
\hat{L}_n(\hat{r}_j)=\dt_{nm}, \quad n,m=0,1,\ldots,M.   
\label{dolagu}   
\eea   
As in the Hermite case, the computation of $\{\omega^r_j\}$ from  
(\ref{lweight}) is   
 not a stable process for very large $N$. However, one can compute   
 $\{\omega^r_j\}$ in a stable way as suggested in the Appendix of  
 \cite{Shen00}.

The memory requirement of this scheme is $O(M)$  and the computational cost   
per time step is a small multiple of $M^2$.   
As for the stability of the TSLP4, we have the following   
\begin{lemma}   
The time-splitting  Laguerre-pseudospectral (TSLP4) method (\ref{schm5})   
is unconditionally stable. More precisely, we have   
\be   
\|\psi^n\|_{l^2}^2=\sum_{j=0}^M \og_j^r  |\psi_j^n|^2   
= \sum_{j=0}^M \og_j^r  |\psi_0(r_j)|^2=\|\psi_0\|_{l^2}^2, \qquad    
n\ge0. \nn  
\ee   
\end{lemma}   
 \begin{proof}  Using (\ref{dolagu}),  the proof is   
   essentially the same   as in  Lemma \ref{tshp4sb}  
for time-splitting Hermite-pseudospectral method. 
\end{proof}   
   
   \subsection{Laguerre-Hermite pseudospectral method for 3-D GPE with   
 cylindrical symmetry}   
In the 3-D case with cylindrical symmetry, i.e. $d=3$ and $\gm_x=\gm_y$   
in (\ref{gpeg1}),   
and $\psi_0(x,y,z)=\psi_0(r,z)$ in (\ref{gpeg2}),    
 the solution of (\ref{gpeg1})-(\ref{gpeg2}) with $d=3$ satisfies    
$\psi(x,y,z,t)=\psi(r,z,t)$. Therefore, Eq. (\ref{step1}) becomes to   
\bea    
\label{gpeg3d}    
&&i\;\pl{\psi(r,z,t)}{t}=B\psi(r,z,t) =-\fl{1}{2}\left[\fl{1}{r}   
\pl{}{r}\left(r\pl{\psi}{r}\right)+ \fl{\p^2 \psi}{\p z^2}   
\right] +\fl{1}{2}\left(\gm_r^2 r^2+\gm_z^2 z^2\right)\psi,\nn\\   
&&  \qquad \qquad\qquad \qquad \qquad \quad 0<r<\ift,\ -\ift<z<\ift,\\   
&&\lim_{r\to\ift} \psi(r,z,t)=0, \quad \lim_{|z|\to\ift} \psi(r,z,t)=0,   
\qquad t\ge0,    
\eea    
where $\gm_r=\gm_x=\gm_y$. The normalization (\ref{normg}) becomes   
\be   
\label{norm7}   
\|\psi(\cdot,t)\|^2 = 2\pi \int_0^\ift \int_{-\ift}^\ift    
|\psi(r,z,t)|^2 r\; dr dz \equiv   
2\pi \int_0^\ift \int_{-\ift}^\ift    
|\psi_0(r,z)|^2 r\; dr dz=1.   
\ee   
We are now in position to present our   
 Laguerre-Hermite pseudospectral    
method for  (\ref{gpeg3d}).   
   
Using the same notations as in previous subsections, we derive from   
(\ref{herm1dd}) and (\ref{lau3d}) that   
\bea   
\label{orgcin}   
\lefteqn{-\fl{1}{2}\left[\fl{1}{r}   
\pl{}{r}\left(r\pl{}{r}\right)+ \fl{\p^2 }{\p z^2}   
\right](L_m(r)\;h_l(z)) +   
\fl{1}{2}\left(\gm_r^2 r^2+\gm_z^2 z^2\right)(L_m(r)\;h_l(z))}\nn\\[2mm]   
&=&\left[-\fl{1}{2r}   
\fl{d}{dr}\left(r\fl{dL_m(r)}{dr}\right) +\fl{1}{2}\gm_r^2 r^2L_m(r)\right]   
h_l(z) +\left[-\fl{1}{2}\fl{d^2h_l(z) }{dz^2}+\fl{1}{2} \gm_z^2 z^2   
h_l(z)\right]L_m(r)\nn\\   
&=& \mu_m^r L_m(r) h_l(z) + \mu_l^z h_l(z) L_m(r)=   
(\mu_m^r+\mu_l^z)L_m(r) h_l(z).   
\eea   
Hence, $\{L_m(r) h_l(z)\}$ are eigenfunctions of $B$ defined in 
(\ref{gpeg3d}).    
   
For a fixed pair $(M,N)$, let    
$X_{MN}=\text{span}\{L_m(r)h_l(z):\,m=0,1,\cdots,M,\;l=0,1,\cdots,N\}$.   
The Laguerre-Hermite spectral method for   
 (\ref{gpeg3d}) is    
 to find $\psi_{MN}(r,z,t)$ $\in   
X_{MN}$,   
 i.e.,    
\be   
\label{expan3dmn}   
\psi_{MN}(r,z,t)=\sum_{m=0}^M \sum_{l=0}^N\tilde{\psi}_{ml}(t) \; L_m(r)\;   
h_l(z)   
\ee   
such that    
\be   
\label{gpeg3dmn}   
\begin{split}   
i\;\pl{\psi_{MN}(r,z,t)}{t}&=B\psi_{MN}(r,z,t)\\   
& =-\fl{1}{2}\left[\fl{1}{r}   
\pl{}{r}\left(r\pl{\psi_{MN}}{r}\right)+ \fl{\p^2 \psi_{MN}}{\p z^2}   
\right] +\fl{1}{2}\left(\gm_r^2 r^2+\gm_z^2 z^2\right)\psi_{MN}.   
\end{split}   
\ee   
Plugging (\ref{expan3dmn})   
into (\ref{gpeg3dmn}), thanks to (\ref{orgcin}), we find that   
\be   
\label{ode3d}   
i\; \fl{d \tilde{\psi}_{ml}(t)}{dt}= \left(\mu_m^r +\mu_l^z\right)    
\tilde{\psi}_{ml}(t),\;m=0,1,\cdots,M,\;l=0,1,\cdots,N.    
\ee   
Hence, the solution for  (\ref{gpeg3dmn}) is given by   
\bea   
\label{sol1d4}   
{\psi}_{MN}(r,z,t)&=&e^{-iB(t-t_s)}{\psi}_{MN}(r,z,t_s)\nn\\ 
&=&\sum_{m=0}^M\sum_{l=0}^N e^{-i(\mu_m^r    
  +\mu_l^z)(t-t_s)} \tilde{\psi}_{ml}    
(t_s)  L_m(r)\;h_l(z),  \qquad t \ge t_s.   
\eea  
   
Let $\psi^n_{jk}$ be the approximation of $\psi(r_j,z_k,t_n)$ and    
$\psi^n$ be the solution vector with components $\psi_{jk}^n$.    
The fourth-order time-splitting Laguerre-Hermite-pseudospectral   
(TSLHP4) method     
for 3-D GPE (\ref{gpeg1}) with cylindrical symmetry   
 is given by    
\bea    
&&\psi^{(1)}_{jk}=e^{-i2w_1\; \Delta t\; \bt_3|\psi_{jk}^{n}|^2}   
\;\psi_{jk}^{n},    \nn\\    
&&\psi_{jk}^{(2)}=\sum_{m=0}^{M} \sum_{l=0}^{N}     
  e^{-i2w_2 \Delta t(\mu_m^r+\mu_l^z)}\;\widehat{(\psi^{(1)})}_{ml}\;   
L_m(r_j)h_l(z_k),      \nn\\     
\label{schm6}     
&&\psi^{(3)}_{jk}=e^{-i2w_3 \; \Delta t\;    
\beta_3|\psi_{jk}^{(2)}|^2}\;\psi_{jk}^{(2)},     
 \nn\\    
&&\psi_{jk}^{(4)}=\sum_{m=0}^{M} \sum_{l=0}^{N}     
  e^{-i2w_4 \Delta t(\mu_m^r+\mu_l^z)}\;\widehat{(\psi^{(3)})}_{ml}\;   
L_m(r_j)h_l(z_k),      \nn\\     
&&\psi^{(5)}_{jk}=e^{-i2w_3 \; \Delta t\;    
\beta_3|\psi_{jk}^{(4)}|^2}\;\psi_{jk}^{(4)},   
\quad j=0,1,\ldots,M, \ k=0,1\ldots,N,    \nn\\   
&&\psi_{jk}^{(6)}=\sum_{m=0}^{M} \sum_{l=0}^{N}     
  e^{-i2w_2 \Delta t(\mu_m^r+\mu_l^z)}\;\widehat{(\psi^{(5)})}_{ml}\;   
L_m(r_j)h_l(z_k),   \nn\\     
&&\psi^{n+1}_{jk}=e^{-i2w_1 \; \Delta t\;   
 \beta_3|\psi_{jk}^{(6)}|^2}\;\psi_{jk}^{(6)};     
\eea    
where    
$\widehat{U}_{ml}$, the  coefficients of   scaled Laguerre-Hermite   
expansion of $U(r,z)$ are computed by the discrete scaled   
Laguerre-Hermite transform   
\be    
\label{Fourc3}    
\widehat{U}_{ml}=   \sum_{j=0}^{M} \sum_{k=0}^{N} \og_j^r\;   
 \og_k^z\; U(r_j,z_k)\; L_m(r_j)   
h_l(z_k),  \ m=0,1,\ldots, M,  \ k=0,1,\ldots,N.   
\ee    
The memory requirement of this scheme is    
$O(MN)$ and the computational cost per time step is    
$O(\max(M^2N,N^2M))$.   
As for the stability of the TSLHP4, we have the following   
\begin{lemma}   
The time-splitting  Laguerre-Hermite pseudospectral (TSLHP4)    
method (\ref{schm6})   
is unconditionally stable. More precisely, we have   
\be   
\|\psi^n\|_{l^2}^2=\sum_{j=0}^M \sum_{k=0}^N \og_j^r \og_k^z %r_j  
|\psi_{jk}^n|^2   
= \sum_{j=0}^M \sum_{k=0}^N \og_j^r \og_k^z  %r_j  
|\psi_0(r_j,z_k)|^2   
=\|\psi_0\|_{l^2}^2, \quad    
n\ge0. \nn  
\ee   
\end{lemma}   
\begin{proof}  Using (\ref{doherm})  and (\ref{dolagu}),  the proof is   
  essentially the same    
  as in  Lemma \ref{tshp4sb}  
for time-splitting Hermite-pseudospectral method. 
\end{proof}

\section{Extension to multi-component BECs}   
\label{exten}  
\setcounter{equation}{0}    
  
The time-splitting Laguerre-Hermite pseudospectral  method,    
introduced  above  for the 3-D GPE with cylindrical symmetry,   
can be  extended to vector Gross-Pitaevskii equations (VGPEs)   
for multi-component BECs \cite{baom}. For simplicity,    
we only present the detailed method for the dynamics of two-component    
BECs.  Consider the dimensionless VGPEs with an external driven filed    
(cf. \cite{baom})   
\bea    
\label{vgpe1}    
&&i\;\pl{\psi(r,z,t)}{t}=-\fl{1}{2}\left[\fl{1}{r}   
\pl{}{r}\left(r\pl{\psi}{r}\right)+ \fl{\p^2 \psi}{\p z^2}   
\right] +\fl{1}{2}\left(\gm_r^2 r^2+\gm_z^2 (z-z_1^0)^2\right)\psi \\   
&&\qquad \qquad \quad +\left(\bt_{11} |\psi|^2 +\bt_{12}|\phi|^2\right)\psi    
+\sqrt{N_2^0/N_1^0} f(t) \phi,\nn\\   
\label{vgpe2}    
&&i\;\pl{\phi(r,z,t)}{t}=-\fl{1}{2}\left[\fl{1}{r}   
\pl{}{r}\left(r\pl{\phi}{r}\right)+ \fl{\p^2 \phi}{\p z^2}   
\right] +\fl{1}{2}\left(\gm_r^2 r^2+\gm_z^2 (z-z_2^0)^2\right)\phi \\   
&&\qquad \qquad \quad  +\left(\bt_{21} 
|\psi|^2 +\bt_{22}|\phi|^2\right)\phi    
+\sqrt{N_1^0/N_2^0} f(t) \psi,  \ 0<r<\ift,\ z\in {\Bbb R}, \nn \\   
&&\lim_{r\to\ift} \psi(r,z,t)=\lim_{r\to\ift} \phi(r,z,t)=0,    
\ \lim_{|z|\to\ift} \psi(r,z,t)=\lim_{|z|\to\ift} \phi(r,z,t)= 0,\\   
&&\psi(r,z,0)=\psi_0(r,z), \qquad \phi(r,z,0)=\phi_0(r,z);    
\eea    
where $z_j^0$ and $N_j^0$ ($j=1,2)$ are the center of   
trapping potential  along    
$z$-axis and  the number of atoms of the     
$j$th component, respectively;   
$\gm_r=\og_r/\og_m$, $\gm_z=\og_z/\og_m$ with    
$\og_r$, $\og_z$ and $\og_m$ are the radial, axial and reference    
frequencies, respectively;   
$\bt_{jl} = 4\pi a_{jl} N_l^0/a_0$ ($j,l=1,2$)   
with the s-wave scattering length  $a_{jl}=a_{lj}$ between the $j$th   
and $l$th component and $a_0=\sqrt{\hbar/m\og_m}$; and   
$f(t) = \Og \cos(\og_d t/\og_m)/\og_m$ with   
$\Og$ and $\og_d$ being the amplitude and frequency of the external   
driven field.    
The wave functions are normalized as    
\bea    
\label{norm8}   
\|\psi_0\|^2&=&2\pi \int_0^\ift \int_{-\ift}^\ift  
|\psi_0(r,z)|^2 r\;drdz=1, \nn \\   
\|\phi_0\|^2&=&2\pi \int_0^\ift \int_{-\ift}^\ift    
|\phi_0(r,z)|^2 r\;drdz=1.   
\eea   
It is easy to show (cf. \cite{baom}) that the total number of atoms is   
conserved     
\bea   
\label{norm9}   
N_1^0\|\psi(\cdot,t)\|^2 +N_2^0 \|\phi(\cdot,t)\|^2    
&=&2\pi \int_0^\ift \int_{-\ift}^\ift\left(|\psi(r,z,t)|^2   
+|\phi(r,z,t)|^2\right)r\;drdz \nn \\   
&\equiv&N_1^0 \|\psi_0\|^2   
+N_2^0 \|\phi_0\|^2 =N_1^0+N_2^0.   
\eea   
Unlike the time-splitting Laguerre-Hermite    
method for 3-D GPE (\ref{gpeg1}) with   
cylindrical symmetry, here we have to split  the VGPEs   
(\ref{vgpe1})-(\ref{vgpe2})     
into  three sub-systems. For example, for a fist-order splitting   
scheme,  we first solve   
\bea    
\label{vgpe11}    
&&i\;\pl{\psi(r,z,t)}{t}=-\fl{1}{2}\left[\fl{1}{r}   
\pl{}{r}\left(r\pl{\psi}{r}\right)+ \fl{\p^2 \psi}{\p z^2}   
\right] +\fl{1}{2}\left(\gm_r^2 r^2+\gm_z^2 z^2\right)\psi, \\   
\label{vgpe21}    
&&i\;\pl{\phi(r,z,t)}{t}=-\fl{1}{2}\left[\fl{1}{r}   
\pl{}{r}\left(r\pl{\phi}{r}\right)+ \fl{\p^2 \phi}{\p z^2}   
\right] +\fl{1}{2}\left(\gm_r^2 r^2+\gm_z^2 z^2\right)\phi,   
\eea   
for the time step of length $\Delta t$, followed by solving    
\bea    
\label{vgpe12}    
&&i\;\pl{\psi(r,z,t)}{t}=\fl{1}{2}\gm_z^2 z_1^0(z_1^0-2z)\psi   
+\left(\bt_{11} |\psi|^2 +\bt_{12}|\phi|^2\right)\psi, \\   
\label{vgpe22}    
&&i\;\pl{\phi(r,z,t)}{t}=\fl{1}{2}\gm_z^2 z_2^0(z_2^0-2z)\phi   
+\left(\bt_{21} |\psi|^2 +\bt_{22}|\phi|^2\right)\phi,   
\eea   
for the same time step, and then by solving    
\bea    
\label{vgpe13}    
&&i\;\pl{\psi(r,z,t)}{t}=\sqrt{N_2^0/N_1^0}\;   
f(t) \;\phi,\\   
\label{vgpe23}    
&&i\;\pl{\phi(r,z,t)}{t}=\sqrt{N_1^0/N_2^0}   
\;f(t)\; \psi,     
\eea    
for the same time step.    
   
The nonlinear    
ODE system (\ref{vgpe12})-(\ref{vgpe22}) leaves    
$|\psi(r,z,t)|$ and $|\phi(r,z,t)|$  invariant in $t$   
and  thus can be integrated {\bf exactly} \cite{baom}.   
The linear ODE system  (\ref{vgpe13})-(\ref{vgpe23})    
can also be integrated {\bf exactly}  by applying a matrix diagonalization   
technique (cf. \cite{baom}). As is shown above,   
(\ref{vgpe11})-(\ref{vgpe21})     
can be discretized in space by Laguerre-Hermite pseudospectral    
method and    
integrated in time  {\bf exactly}.    
   
Let $\psi^n_{jk}$ and $\phi^n_{jk}$ be the approximations of    
$\psi(r_j,z_k,t_n)$  and $\phi(r_j,z_k,t_n)$, respectively,  and    
$\psi^n$  and $\phi^n$ be the solution vectors with components    
$\psi_{jk}^n$ and $\phi_{jk}^n$, respectively. Although it is not   
clear how to construct a fourth-order time splitting schemes with   
three sub-systems, a second-order scheme can be   
easily constructed    
using the Strang splitting (cf. \cite{Stra68}). More precisely,   
from time $t=t_n$ to    
$t=t_{n+1}$, we proceed as follows:    
\bea    
&&\psi_{jk}^{(1)}=\sum_{m=0}^{M} \sum_{l=0}^{N}     
  e^{-i(\mu_m^r+\mu_l^z)\Delta t/2}\;\widehat{(\psi^{(1)})}_{ml}\;L_m(r_j)   
h_l(z_k),      \nn\\     
&&\phi_{jk}^{(1)}=\sum_{m=0}^{M} \sum_{l=0}^{N}     
  e^{-i(\mu_m^r+\mu_l^z)\Delta t/2}\;\widehat{(\phi^{(1)})}_{ml}\;L_m(r_j)   
h_l(z_k),      \nn\\     
&&\psi_{jk}^{(2)}=e^{-i[\gm_z^2 z_1^0(z_1^0-2z_k)/2   
+(\bt_{11} |\psi_{jk}^{(1)}|^2 +\bt_{12}|\phi_{jk}^{(1)}|^2)]\Delta t/2}   
\psi_{jk}^{(1)}, \nn \\   
&&\phi_{jk}^{(2)}=e^{-i[\gm_z^2 z_2^0(z_2^0-2z_k)/2   
+(\bt_{21} |\psi_{jk}^{(1)}|^2 +\bt_{22}|\phi_{jk}^{(1)}|^2)]\Delta t/2}   
\phi_{jk}^{(1)}, \nn \\   
&&\psi_{jk}^{(3)}=\cos(g(t_{n+1},t_n)) \psi_{jk}^{(2)}
 -i\sin(g(t_{n+1},t_n))   
\sqrt{N_2^0/N_1^0}\phi_{jk}^{(2)}, \nn\\   
&&\phi_{jk}^{(3)}=-i \sin(g(t_{n+1},t_n))   
\sqrt{N_1^0/N_2^0}\psi_{jk}^{(2)} +\cos(g(t_{n+1},t_n)) \phi_{jk}^{(2)},  
 \nn\\   
&&\psi_{jk}^{(4)}=e^{-i[\gm_z^2 z_1^0(z_1^0-2)/2   
+(\bt_{11} |\psi_{jk}^{(3)}|^2 +\bt_{12}|\phi_{jk}^{(3)}|^2)]\Delta t/2}   
\psi_{jk}^{(3)}, \nn \\   
&&\phi_{jk}^{(4)}=e^{-i[\gm_z^2 z_2^0(z_2^0-2)/2   
+(\bt_{21} |\psi_{jk}^{(3)}|^2 +\bt_{22}|\phi_{jk}^{(3)}|^2)]\Delta t/2}   
\phi_{jk}^{(3)}, \quad 0\le j\le M, \ 0\le k\le N,   \nn \\   
&&\psi_{jk}^{n+1}=\sum_{m=0}^{M} \sum_{l=0}^{N}     
  e^{-i(\mu_m^r+\mu_l^z)\Delta t/2}\;\widehat{(\psi^{(4)})}_{ml}\;L_m(r_j)   
h_l(z_k),      \nn\\     
\label{multiBEC}   
&&\phi_{jk}^{n+1}=\sum_{m=0}^{M} \sum_{l=0}^{N}     
  e^{-i(\mu_m^r+\mu_l^z)\Delta t/2}\;\widehat{(\phi^{(4)})}_{ml}\;L_m(r_j)   
h_l(z_k),      
\eea    
where    
\[ g(t,t_n) = \int_{t_n}^t f(s) \; ds = {\Og}{\og_d}   
\left[\sin(\og_d t/\og_m) -\sin(\og_d t_n /\og_m)\right]. \]   
Note that the only time discretization error of    
 this scheme is the splitting    
error, which is of second  order in $\Delta t$.   
The scheme  is explicit,   
  spectral accurate in space and second order accurate in time.   
The memory requirement of this method is    
$O(MN)$ and the computational cost per time step is    
$O(\max(M^2N,MN^2))$.   
As for the stability, we can prove as in \cite{baom} the  
following lemma, which   
shows that the total number of atoms is conserved in the discretized level.   
\begin{lemma}   
The time-splitting  Laguerre-Hermite-pseudospectral    
method (\ref{multiBEC}) for multi-component BEC    
is unconditionally stable. More precisely, we have   
\be   
N_1^0\|\psi^n\|_{l^2}^2+N_2^0\|\phi^n\|_{l^2}^2%&=&   
%\sum_{j=0}^M \sum_{s=0}^N \og_j^r \og_s^z   
%\left(N_1^0|\psi_{js}^n|^2+N_2^0|\phi_{js}^n|^2\right) \nn \\   
%= \sum_{j=0}^M \sum_{s=0}^N \og_j^r \og_s^z \left(N_1^0|\psi_0(r_j,z_s)   
%+N_2^0|\phi_0(r_j,z_s)\right),\qquad    
=N_1^0\|\psi_0\|_{l^2}^2+N_2^0\|\phi_0\|_{l^2}^2, \quad    
n\ge0. \nn  
\ee   
\end{lemma}

\section{Numerical results}\label{sne}    
\setcounter{equation}{0}    
    
We now present some numerical results  by using the numerical methods   
introduced in section \ref{smethod}. To quantify the numerical results,    
we define the condensate width along $r$- and $z$-axis as   
\[\sg_\ap^2 = \int_{{\Bbb R}^d} \ap^2 |\psi(\bx,t)|\; d\bx,   
\quad \ap=x,y,z, \qquad \sg_r^2 = \sg_x^2 +\sg_y^2.\]

\begin{figure}[htb]    
\centerline{a).\psfig{figure=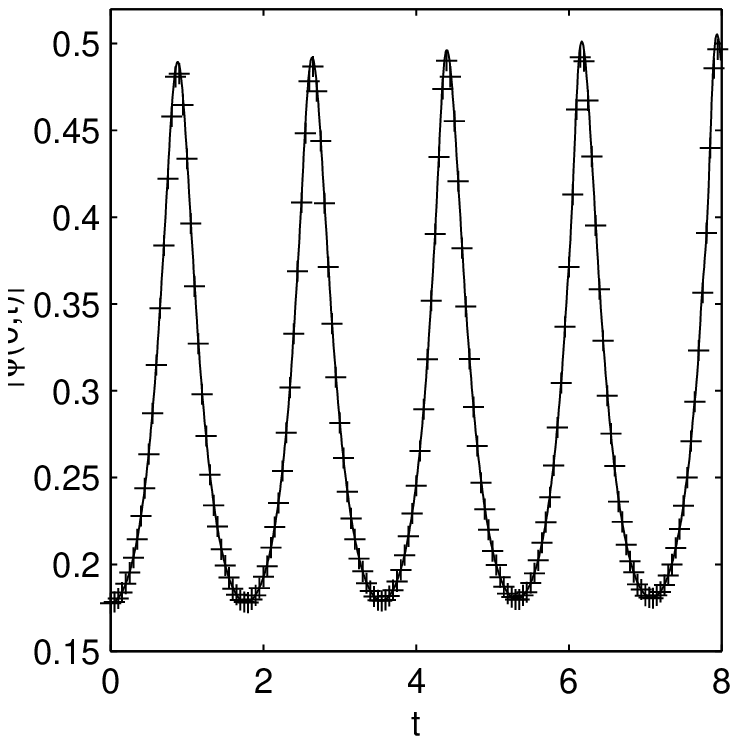,height=6cm,width=6cm,angle=0}   
\quad  b).\psfig{figure=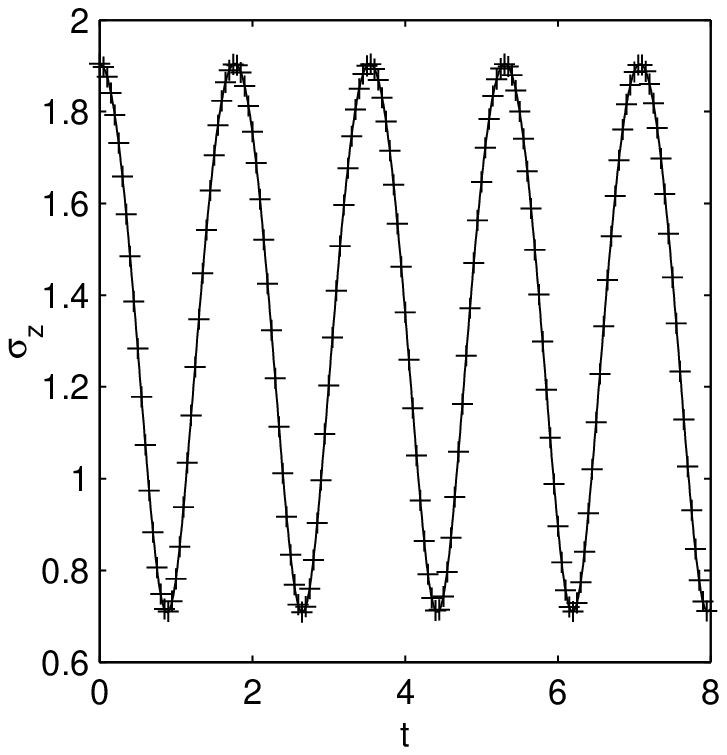,height=6cm,width=6cm,angle=0} }     
    
Figure 1:  Evolution of central density and condensate width in Example 1.    
`---': `exact solutions' obtained by TSSP \cite{Bao3} with    
$1025$ grid points over an interval   
$[-12,12]$;  `+ + + ': Numerical results by TSHP4  (\ref{schm4})    
with $31$ grid points on the whole $z$-axis.   
a). Central density $|\psi(0,t)|^2$; b). condensate width $\sg_z$.   
\end{figure}    
    
 {\bf Example 1.} The 1-D Gross-Pitaevskii equation:  We choose $d=1$,   
 $\gm_z=2$, $\bt_1=50$ in (\ref{gpeg1}).     
 The initial data    
$\psi_0(z)$ is chosen as the ground state of the 1-D  GPE (\ref{gpeg1})   
with $d=1$, $\gm_z=1$ and $\bt_1=50$ \cite{BD,Bao}.   
This corresponds to an experimental setup where  
initially the condensate is    
assumed to be in its ground state, and the trap frequency is double   
 at $t=0$.    
 We solve this problem   
by using (\ref{schm4}) with $N=31$  and time step    
$k=0.001$. Figure 1 plots the condensate width and central density   
$|\psi(0,t)|^2$ as functions of time.   
Our numerical experiments also show that the scheme (\ref{schm4})   
with $N=31$ gives similar numerical results as the TSSP    
method \cite{Bao3,Bao4} for this example with $129$ grid points over    
the interval $[-12,12]$ and time step $k=0.001$.

\bigskip    
    
 {\bf Example 2.} The 2-D  Gross-Pitaevskii equation with radial symmetry:   
we choose $d=2$, $\gm_r=\gm_x=\gm_y=2$, $\bt_2=50$  in   
(\ref{gpeg1}). The initial data     
$\psi_0(r)$ is chosen as the ground state of the 2-D GPE (\ref{gpeg1})   
with $d=2$, $\gm_r=\gm_x=\gm_y=1$ and $\bt_2=50$ \cite{BD,Bao}.    
Again this corresponds to an experimental    
setup where initially the condensate is    
assumed to be in its ground state, and  the trap frequency is doubled    
 at $t=0$.    
We solve this problem   
by using (\ref{schm5}) with $M=30$  and time step    
$k=0.001$. Figure 2 plots the condensate width and central density   
$|\psi(0,t)|^2$ as functions of time.   
Our numerical experiments also show that the scheme (\ref{schm5})   
with $M=30$ gives similar numerical results as the TSSP    
method \cite{Bao3,Bao4} for this example with $129^2$ grid points over    
the box $[-8,8]^2$ and time step $k=0.001$.

\begin{figure}[htb]    
\centerline{a).\psfig{figure=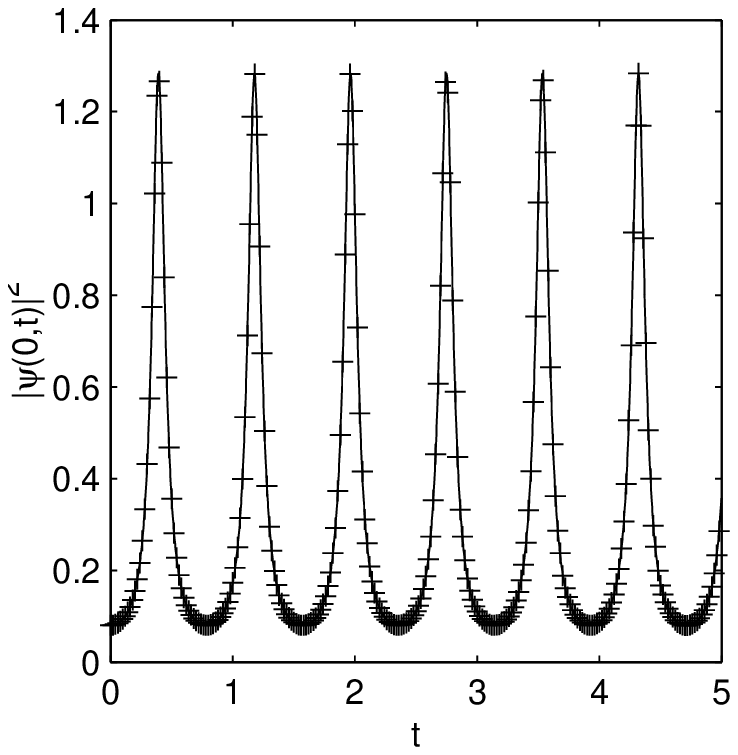,height=6cm,width=6cm,angle=0}   
\qquad  b).\psfig{figure=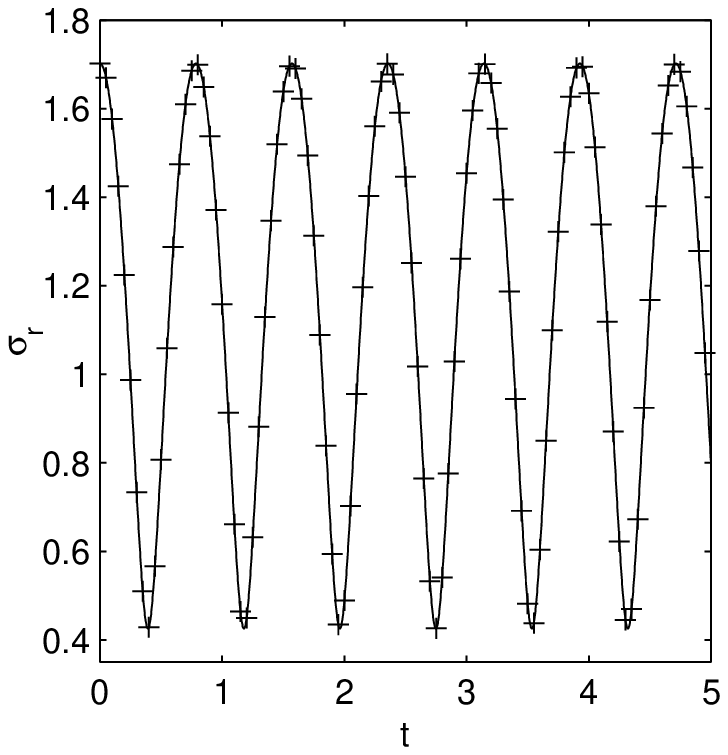,height=6cm,width=6cm,angle=0} }     
    
Figure 2:  Evolution of central density and condensate width in Example 2.    
`---': `exact solutions' obtained by TSSP \cite{Bao3} with    
$1029^2$ grid points over a box   
$[-8,8]^2$;  `+ + + ': Numerical results by TSLP4 (\ref{schm5})    
with $30$ grid points on the semi-infinite interval $[0,\ift)$.   
a). Central density $|\psi(0,t)|^2$; b). condensate width $\sg_r$.   
\end{figure}    
    
 {\bf Example 3.} The 3-D  Gross-Pitaevskii equation with cylindrical    
symmetry:   
we choose $d=3$, $\gm_r=\gm_x=\gm_y=4$, $\gm_z=1$ and   
 $\bt_3=100$  in (\ref{gpeg1}). The initial data    
$\psi_0(r,z)$ is chosen as the ground state of the 3-D GPE (\ref{gpeg1})   
with $d=3$, $\gm_r=\gm_x=\gm_y=1$, $\gm_z=4$  and $\bt_3=100$ \cite{BD,Bao}.   
This corresponds to an experimental setup where initially the condensate is    
assumed to be in its ground state, and  at $t=0$,    
we increase the radial  frequency four times and decrease the axial   
frequency to its quarter.   
We solve this problem   
by using (\ref{schm6}) with $M=60$ and $N=61$   and time step    
$k=0.001$. Figure 3 plots the condensate widths and central density   
$|\psi(0,0,t)|^2$ as functions of time.

The numerical results for these three examples clearly indicated that   
 our new methods are very  efficient and accurate.

\begin{figure}[htb]    
\centerline{\psfig{figure=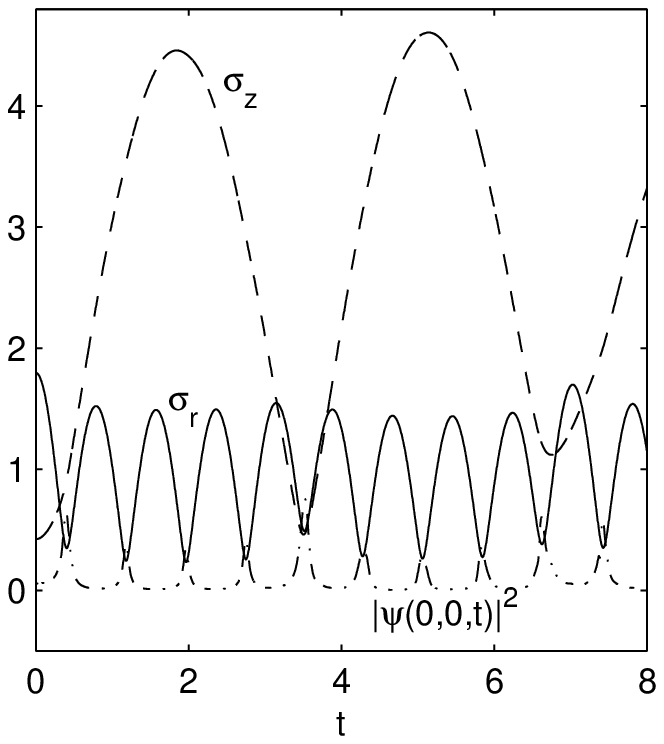,height=6cm,width=12cm,angle=0} }     
    
Figure 3: Evolution of central density and condensate width   
 in Example 3 by the TSLHP4 (\ref{schm6}) .    
\end{figure}

 {\bf Example 4.} The 3-D  vector Gross-Pitaevskii equations    
with cylindrical    
symmetry for two-component BECs:   
we take,  in (\ref{vgpe1}) and (\ref{vgpe2}),   
$m=1.44\tm10^{-25}\;[kg]$, $a_{12}=a_{21}=55.3\AA=5.53\;[nm]$,    
$a_{11}=1.03a_{12}=5.6959\;[nm]$,  $a_{22}=0.97a_{12}=5.3641\;[nm]$,   
 $\og_{z}=47\tm 2\pi\;[1/s]$,    
$\og_m=\og_{r}=\og_{z}/\sqrt{8}$,     
$N_1^0=N_2^0=500,000$, $\Og = 65\tm2\pi\;[1/s]$,    
$\og_d=6.5\tm 2\pi\;[1/s]$.   
A simple computation shows     
$a_0=0.2643\tm 10^{-5}\;[m]$, $\bt_{11}=0.02708165N_1^0$,    
$\bt_{12}=0.02629286N_2^0$, $\bt_{21}=0.02629286N_1^0$,    
$\bt_{22}=0.02550407N_2^0$.  The initial data    
$\psi_0(r,z)$ and $\phi_0(r,z)$ are chosen as the    
ground state of the 3-D VGPEs (\ref{vgpe1}) and (\ref{vgpe2}), and   
we set $f(t)\equiv0$ \cite{baom}.   
We solve this problem   
by using (\ref{multiBEC}) with $M=100$ and $N=201$ and time step    
$k=0.00025$. Figure 4 displays the time evolution of the   
density functions for the two components with different    
trapping centers. The results are similar as those obtained    
in \cite{baom} by a TSSP method with  a much refined grid.   
   
\begin{figure}   
 \centerline{a). \psfig{figure=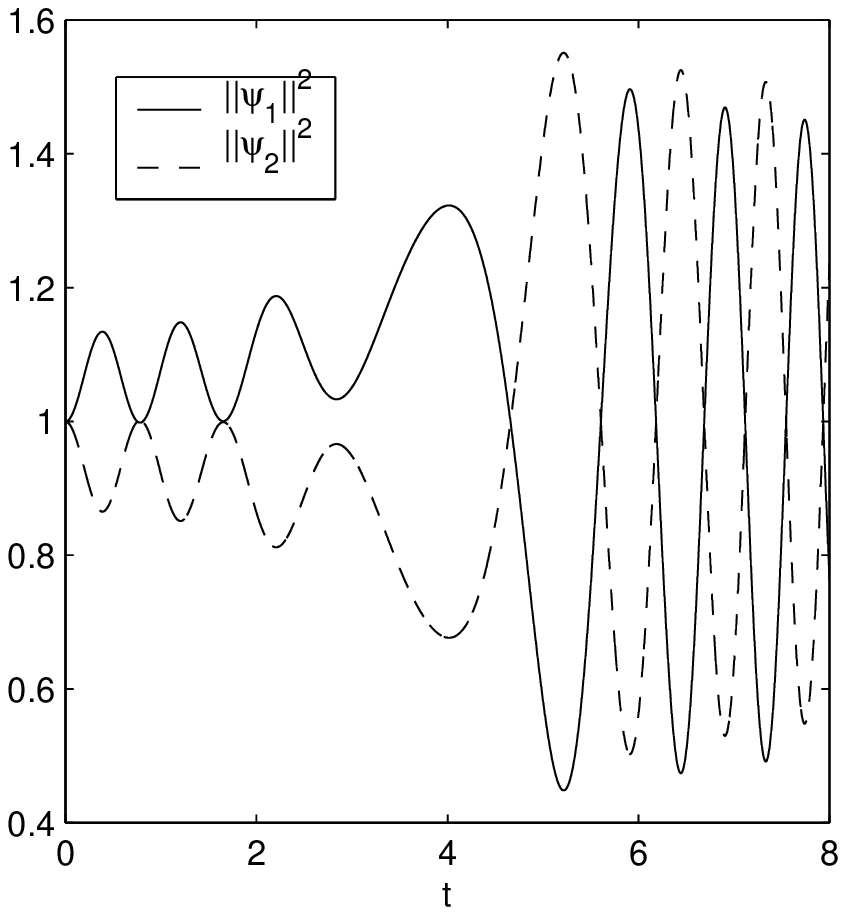,height=6cm,width=12cm,angle=0} }     
\centerline{b). \psfig{figure=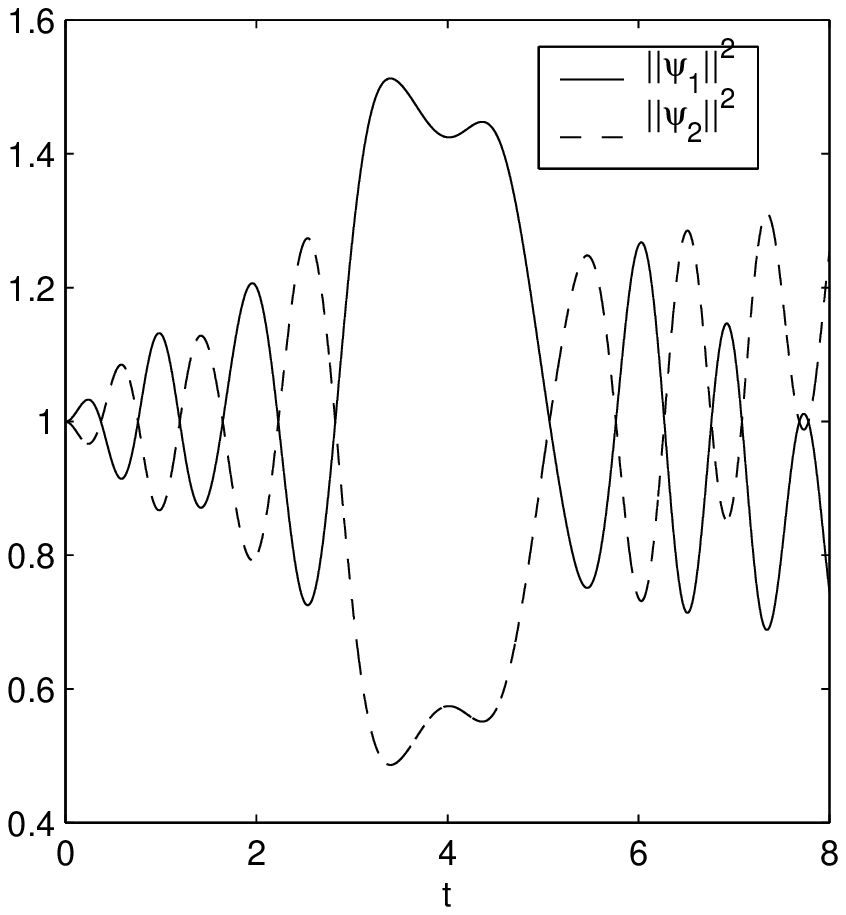,height=6cm,width=12cm,angle=0} }     
 \centerline{c). \psfig{figure=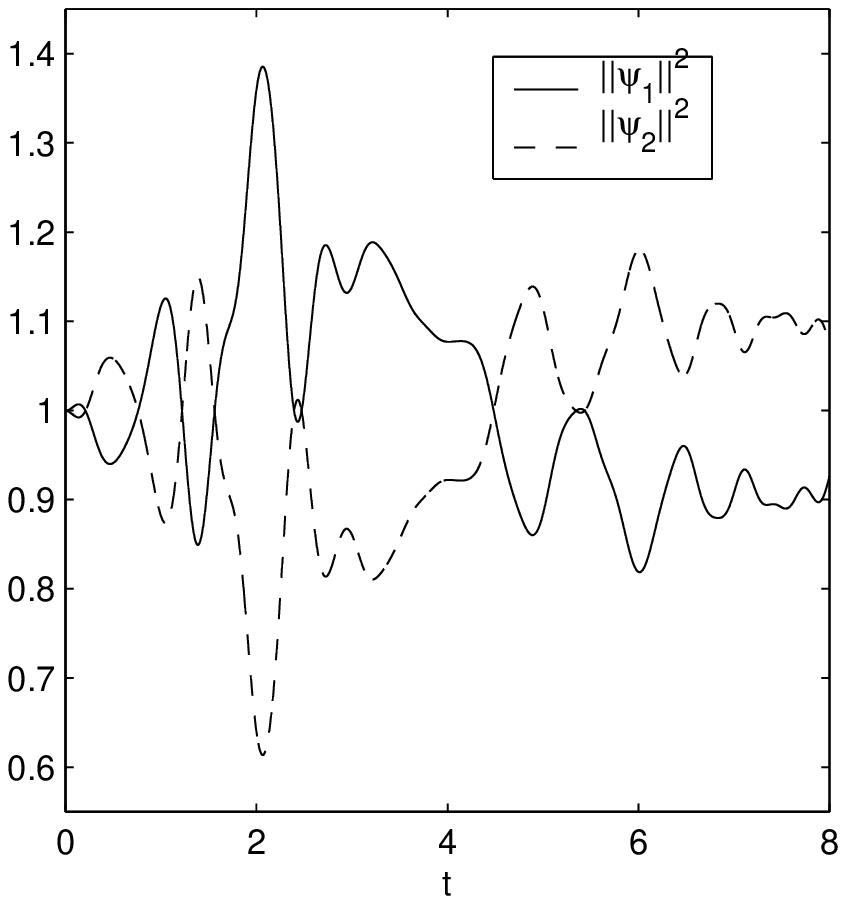,height=6cm,width=12cm,angle=0} }     
    
Figure 4: Time evolution of the density functions   
for the two-component BECs in Example 4.    
 a). $z_1^0=z_2^0=0$, b). $z_1^0=-z_2^0=0.15$, c). $z_1^0=-z_2^0=0.4$.   
\end{figure}    
   
>From Fig. 4, we can see that    
the general form of time evolution on the number of particles   
in the two components is similar for different distances    
between the two trapping potential centers.   
When $z_1^0=z_2^0=0$, the number of particles in the   
second component, i.e. $N_2^0 \|\phi\|^2$, decreases, reaching   
its bottom, oscillates and then attains its maximum at    
around  $t=5.2$. The number of particles   
in the second component at its maximum is approximately 55\% bigger   
than its initial value at time $t=0$.   
 (cf. Fig. 4a).  The pattern for  $N_1^0 \|\psi\|^2$   
is exactly the  opposite of that for $N_2^0 \|\phi\|^2$   
(cf. Fig. 4a). This antisymmetry is   
 due to the fact that the total number of particles in the    
two components are conserved. When $z_1^0-z_2^0>0$  becomes larger,   
i.e. initially the density functions for the two    
components are separated further, the earlier    
the number of particles attains its absolute peak (cf. Fig. 4b\&c),   
the smaller the maximum of the peak is. In fact, when    
$z_1^0-z_2^0=0.3$ (resp. 0.8), at around   $t=3.4$ (resp. 2.05), the   
number of particles    
in the first component attains its maximum which is   
 approximately 52\% (resp. 38.5\%) bigger  than its initial value at   
 time $t=0$.

\section{Concluding remarks}\label{sc}    
\setcounter{equation}{0}    
    
   We developed a new efficient  fourth-order     
time-splitting Laguerre-Hermite pseudospectral method for    
3-D Gross-Pitaevskii equation  with cylindrical symmetry    
for Bose-Einstein condensates. The new method takes advantage of the   
cylindrical symmetry  so only   
an effective 2-D problem is solved numerically. The new method is based on   
appropriately scaled Laguerre-Hermite  functions and  a   
fourth-order symplectic integrator. Hence, it is   
 spectrally accurate in space, fourth-order accurate in time,   
 explicit,  unconditionally stable, time reversible  and time transverse   
invariant.    
   
When Compared with  the time-splitting   
sine-spectral method in \cite{Bao3,Bao4,baom}  and the   
Crank-Nicolson finite difference method in \cite{Rup,Adh1}, the new   
method enjoys  two important advantages:   
 (i) there is no need to truncate the original whole space    
into a bounded computational domain for which an artificial boundary   
condition (which often erodes  the accuracy) is needed; (ii) it solves   
an effective    
2-D problem    
instead of the original 3-D equations.    
Thus, the new method is very  accurate and efficient, particularly  in   
term of memory requirement.   
Therefore,   
it is extremely suitable for 3-D GPE with cylindrical    
symmetry which is the most frequently setup in BEC experiments.    
We plan to apply this powerful numerical    
 method to study physically more complex systems like multi-component  
 BECs, vortex states and dynamics in BECs.

%\bibliographystyle{plain}   
%\bibliography{../../bibtex/bib}   
%\end{document}   

\end{document}